\documentclass{jfm}

\usepackage{natbib}
\usepackage{amsfonts,amsbsy,amsmath}
\usepackage{graphicx}
\usepackage{color} 
\usepackage{natbib} 
\usepackage{tikz}
\usetikzlibrary{intersections,shapes.arrows}

\usepackage{etoolbox}
\usepackage{upgreek}
\usepackage{lineno}

\newcommand*\patchAmsMathEnvironmentForLineno[1]{%
  \expandafter\let\csname old#1\expandafter\endcsname\csname #1\endcsname
  \expandafter\let\csname oldend#1\expandafter\endcsname\csname end#1\endcsname
  \renewenvironment{#1}%
     {\linenomath\csname old#1\endcsname}%
     {\csname oldend#1\endcsname\endlinenomath}}%
\newcommand*\patchBothAmsMathEnvironmentsForLineno[1]{%
  \patchAmsMathEnvironmentForLineno{#1}%
  \patchAmsMathEnvironmentForLineno{#1*}}%
\AtBeginDocument{%
\patchBothAmsMathEnvironmentsForLineno{equation}%
\patchBothAmsMathEnvironmentsForLineno{align}%
\patchBothAmsMathEnvironmentsForLineno{flalign}%
\patchBothAmsMathEnvironmentsForLineno{alignat}%
\patchBothAmsMathEnvironmentsForLineno{gather}%
\patchBothAmsMathEnvironmentsForLineno{multline}%
}

\DeclareMathAlphabet{\mathsfit}{T1}{\sfdefault}{\mddefault}{\sldefault}

\makeatletter
\newcommand{\specialnumber}[1]{%
	\def\tagform@##1{\maketag@@@{(\ignorespaces##1\unskip\@@italiccorr#1)}}}
\newcommand{\specialeqref}[2]{\begingroup
	\def\tagform@##1{\maketag@@@{(\ignorespaces##1\unskip\@@italiccorr#2)}}%
	\eqref{#1}\endgroup}

\def\e{\mathrm{e}}

\def\d{\mathrm{d}}

\def\beq{\begin{equation}}
\def\eeq{\end{equation}}
\def\bx{\mathbf{x}}
\def\eps{\varepsilon}

\def\bx{\boldsymbol{x}}
\def\bk{\boldsymbol{k}}
\def\by{\boldsymbol{y}}
\def\bu{\boldsymbol{u}}
\def\bK{\boldsymbol{K}}

\def\bX{\boldsymbol{X}}

\def\be{\boldsymbol{e}}
\def\bU{\boldsymbol{U}}

\def\P{\mathsf{P}}

\def\D{\mathsfit{D}}
\def\DD{\mathsfbi{D}}

\newcommand{\bs}[1]{\boldsymbol{#1}}

\def\bc{\bs{c}}

\newcommand{\av}[1]{\langle #1 \rangle}
\newcommand{\dpar}[2]{\frac{\partial #1}{\partial #2}}

\newcommand{\dt}[2]{\frac{\mathrm{d} #1}{\mathrm{d} #2}}

\newcommand{\eqn}[1]{(\ref{eqn:#1})}
\newcommand{\lab}[1]{\label{eqn:#1}}
\newcommand{\inter}[1]{\quad \textrm{#1} \quad}

\definecolor{armygreen}{rgb}{0.29, 0.33, 0.13}

\newrobustcmd*{\smallcircle}[1]{\tikz{\filldraw[draw=#1,fill=#1] (0,0) circle [radius=0.035cm];}}
\newrobustcmd*{\medcircle}[1]{\tikz{\filldraw[draw=#1,fill=#1] (0,0) circle [radius=0.05cm];}}
\newrobustcmd*{\bigcircle}[1]{\tikz{\filldraw[draw=#1,fill=#1] (0,0) circle [radius=0.06cm];}}
\newrobustcmd*{\myredline}{\raisebox{2pt}{\tikz{\draw[red,solid,line width=0.8pt](0,0) -- (3mm,0);}}}
\newrobustcmd*{\myblueline}{\raisebox{2pt}{\tikz{\draw[blue,solid,line width=0.8pt](0,0) -- (3mm,0);}}}
\newrobustcmd*{\myorangeline}{\raisebox{2pt}{\tikz{\draw[orange,solid,line width=0.8pt](0,0) -- (3mm,0);}}}
\newrobustcmd*{\myblackline}{\raisebox{2pt}{\tikz{\draw[black,solid,line width=0.8pt](0,0) -- (3mm,0);}}}

\def\XXint#1#2#3{{\setbox0=\hbox{$#1{#2#3}{\int}$}
\bcenter{\hbox{$#2#3$}}\kern-.5\wd0}}

\title[Diffusion of inertia-gravity waves]{Diffusion of inertia-gravity waves by geostrophic turbulence}
\author[H. A. Kafiabad, M. A. C. Savva \& J. Vanneste]{Hossein A. Kafiabad, Miles A. C. Savva and Jacques Vanneste} 
\affiliation{School of Mathematics and Maxwell Institute for Mathematical Sciences, \\
University of Edinburgh, Edinburgh EH9 3FD, UK}

\begin{document}

\maketitle

\begin{abstract}
The scattering of inertia-gravity waves by large-scale geostrophic turbulence in a rapidly rotating, strongly stratified fluid leads to the diffusion of wave energy on the constant-frequency cone in wavenumber space. We derive the corresponding diffusion equation and relate its diffusivity to the wave characteristics and the energy spectrum of the turbulent flow. We check the predictions of this equation against numerical simulations of the three-dimensional Boussinesq equations in initial-value and forced scenarios with horizontally isotropic wave and flow fields. In the forced case, wavenumber diffusion results in a $k^{-2}$ wave energy spectrum consistent with as-yet-unexplained features of observed atmospheric and oceanic spectra.

\end{abstract}

\section{Introduction}

The dynamics of rotating stratified fluids, most notably the atmosphere and ocean, is characterised by the coexistence of vortical flow and inertia-gravity waves (IGWs). These evolve independently at a linear level but interact to an increasing degree as flow strength and wave amplitude increase. In the weakly nonlinear regime, corresponding to small Rossby and/or Froude numbers, the vortical flow has a `catalytic' role, enabling the scattering of energy between IGWs through resonant triad interactions while remaining unaffected \citep{lelo-rile,bart95,ward-dewa}. The qualitative impact of this catalytic interaction has been considered: an isotropic turbulent flow causes the isotropisation of the IGW field \citep{lelo-rile,savv-vann} and a cascade of wave energy to small scales \citep{bart95,wait-bart06a}. 

Here we provide a quantitative description by deriving a simplified model for the dynamics of IGWs in a low-Rossby-number, homogeneous and horizontally isotropic turbulent flow in geostrophic balance. The derivation (in \S\ref{sec:diffusion}) assumes linear IGWs with small spatial scales relative to the flow. It yields a diffusion equation that captures the spreading of IGWs in wavenumber space or, more precisely, on a cone in this space corresponding to fixed-frequency IGWs. The diffusivity components associated with radial and angular diffusion on the cone are obtained in closed forms involving the IGW parameters and the energy spectrum of the geostrophic flow. Early versions were proposed by \citet{mull-olbe} and \citet{mull76,mull77}. 

We solve the diffusion equation for an initial-value problem (\S\ref{sec:ivp}) and a steady forced problem (\S\ref{sec:forced}), assuming horizontally isotropic IGW fields, and we test the results against numerical simulations of the three-dimensional Boussinesq equations, finding good agreement in both cases. With forcing, the diffusion equation predicts a constant-flux, steady energy spectrum scaling with wavenumber as $k^{-2}$ which is realised numerically. 

Our results are relevant to important open questions about the nature of submesoscale motion in the ocean and mesoscale motion in the atmosphere. Recent data analyses by \citet{buhl-et-al14} and \citet{cali-et-al14,cali-et-al16} led them to hypothesise these motions are dominated by almost linear IGWs. The prediction of a $k^{-2}$ spectrum lends  support to this hypothesis by identifying a robust mechanism -- diffusion by turbulence -- that produces a spectrum consistent with observations (see \S\ref{sec:forced}). 
As for the initial-value predictions, they provide estimates for the time scale of the scale cascade of the IGWs that leads ultimately to their dissipation.

%
%
%
%

\section{Diffusion in wavenumber space} \label{sec:diffusion}

We consider the dynamics of IGWs propagating in a turbulent vortical flow of much larger spatial scale so that the WKB approximation applies. The distribution of wave energy in the $(\bx,\bk)$ phase space is then governed by the conservation 
\beq
\partial_t a + \nabla_{\bk}   \Omega \cdot \nabla_{\bx} a - \nabla_{\bx} \Omega \cdot \nabla_{\bk}   a = 0
\lab{action}
\eeq
of the action density $a(\bx,\bk,t)$. Here $\Omega = \omega + \bU \cdot \bk$ is the frequency, which sums the intrinsic frequency
\beq
\omega = \sqrt{f^2 \cos^2 \theta + N^2 \sin^2 \theta},
\lab{frequency}
\eeq
where $f<N$ are the Coriolis and buoyancy frequencies and $\theta$ is the angle between the wavevector $\bk$ and the vertical, and the Doppler shift $\bU \cdot \bk$, where $\bU=\bU(\bx,t)$ is the vortical flow velocity.
Assuming that the flow is (i) weak enough that $\omega \gg \bU \cdot \bk$, (ii) evolving on a time scale much longer than $\omega^{-1}$, and (iii) well modelled by a homogeneous and stationary random field, we can approximate \eqn{action} by
\beq
\partial_t a + \bc \cdot \nabla_{\bx} a = \nabla_{\bk} \cdot \left(\DD \cdot \nabla_{\bk} a \right),
\lab{diffusion}
\eeq
where $\bc = \nabla_{\bk}\omega$ is the intrinsic group velocity and $\DD$ a $\bk$-dependent diffusivity tensor (see Appendix \ref{app:diffusionEquation} for a derivation). The right-hand side of \eqn{diffusion} captures the scattering of wave action that results from  small-but-sustained random Doppler shifting by the flow; in the regime considered, this naturally leads to diffusion in $\bk$-space. In Cartesian coordinates, the diffusivity tensor takes the form
\beq
\D_{ij}(\bk) = - \frac{1}{2} k_m k_n \int_{-\infty}^\infty \dpar{^2 \Pi_{mn}}{x_i \partial x_j}(\bc(\bk)s) \, \d s,
\lab{diffusivity}
\eeq
where $\Pi_{mn}(\bx) = \av{U_m(\by + \bx) U_n(\by)}$ is the velocity correlation tensor, with $\av{\cdot}$ denoting ensemble average, and summation over repeated indices is implied. 
An analogous expression was obtained by \citet{mcco-bret} in the context of wave--wave interactions in the induced-diffusion regime \citep[see] [\S5, for a review]{mull-et-al}. \citet{mull-olbe} and \citet{mull76,mull77} discussed a flow-induced diffusivity that differs from \eqn{diffusivity} to account heuristically for wave--wave interactions and dissipation. 

 \begin{figure}
    \centering
    \begin{minipage}{.54\linewidth}
        \centering
        \includegraphics[width=1\linewidth]{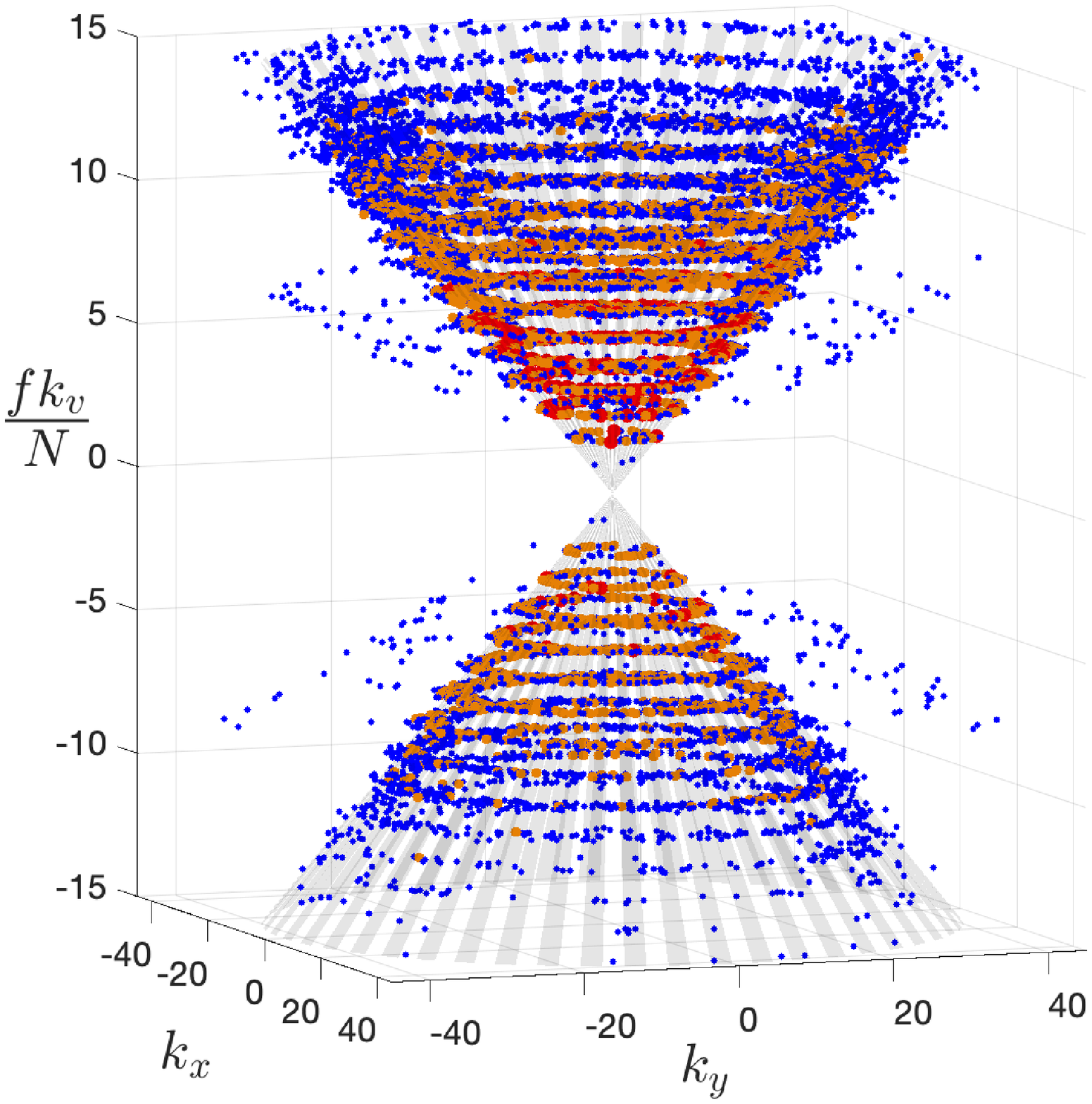}
    \end{minipage}%
    \begin{minipage}{.46\linewidth}
    \centering
        \centering
        \includegraphics[width=1\linewidth]{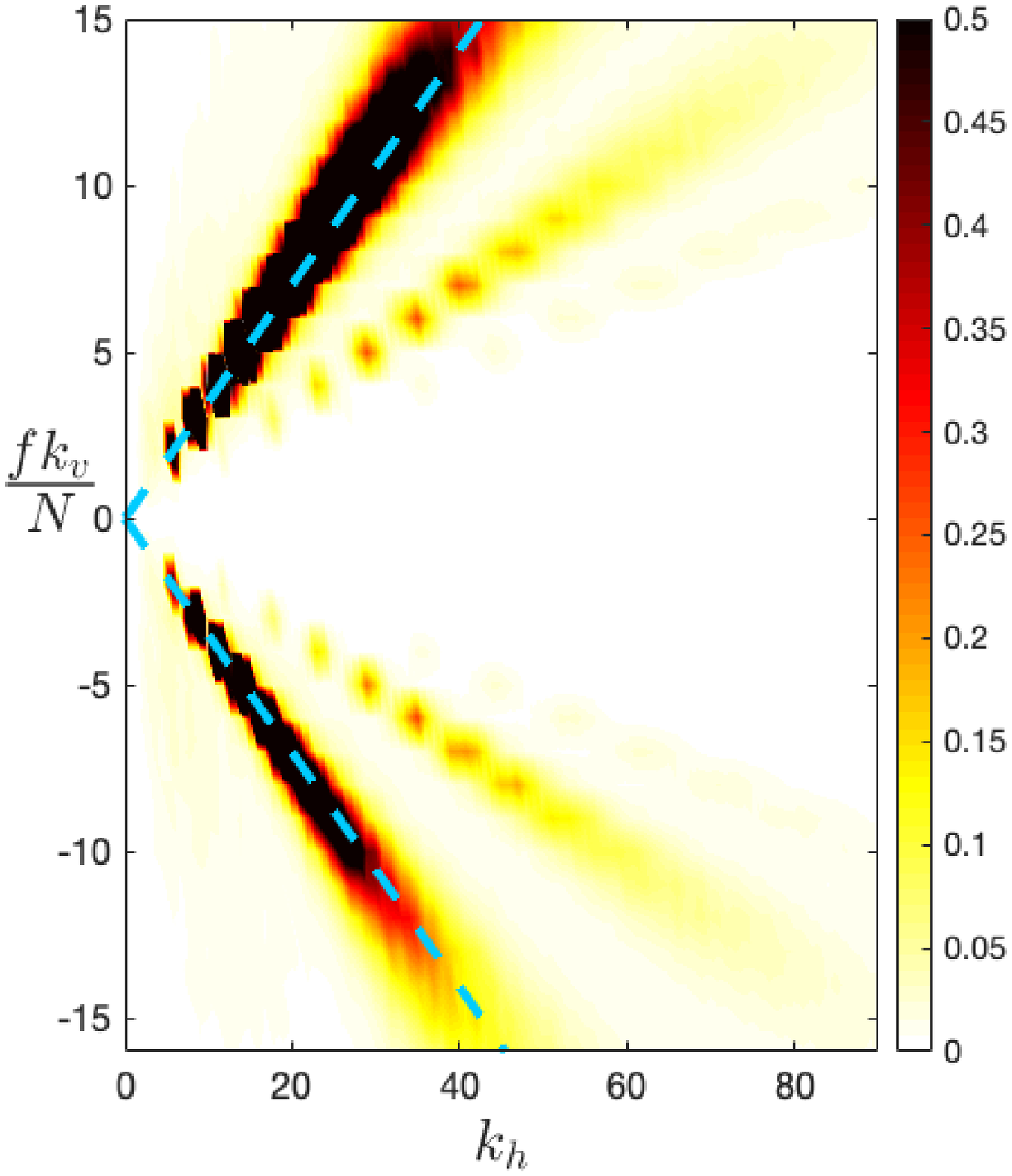}
    \end{minipage}%
\caption{Left: IGW energy density $e(\bk)$ in $\bk$-space at $t=389 f^{-1}$ for the initial-value simulation of \S\ref{sec:ivp} with $\omega=3f$ and $\mathrm{Ro}=0.057$.  \thinspace \bigcircle{red} represents the wave modes with $e(\bk) /e_{\textrm{max}} > 0.1$ ($e_{max}$ the maximum energy density),  \medcircle{orange} $ 0.01 < e(\bk) /e_{\textrm{max}} < 0.1$, and \smallcircle{blue} $ 0.03 < e(\bk) /e_{\textrm{max}} < 0.01$. Right: projection of $e(\bk)$ onto the $(k_\mathrm{h},k_\mathrm{v})$-plane. The constant-frequency cone defined by \eqn{frequency} is indicated by grey stripes on the left and dashed lines on the right.}
\label{fig:cones}
\end{figure}
A key property of  \eqn{diffusivity} is that $\DD(\bk) \cdot \bc(\bk)=0$ since 
\beq
\D_{ij}(\bk) \cdot c_j(\bk)=- \frac{1}{2} k_m k_n \int_{-\infty}^\infty \dt{}{s} \left( \dpar{\Pi_{mn}}{x_i}(\bc(\bk)s) \right) \, \d s = 0.
\eeq
Thus there is no diffusion in the direction of the group velocity $\bc$. Since $\bc$ is perpendicular to constant-frequency surfaces,  for the IGW dispersion relation \eqn{frequency} diffusion is restricted to the cones $\theta=\mathrm{const}$, see Fig.\ \ref{fig:cones}. This is because diffusion in $\bk$-space stems from
 resonant-triad interactions between two IGWs and one vortical mode (also termed balanced mode) associated with the flow. The flow is treated as a zero-frequency mode because it evolves slowly compared with $\omega^{-1}$, so the resonance condition implies that the interacting IGWs have the same frequency. The restriction to a single frequency means that wave action and wave energy only differ by a constant multiple and can be identified with one another. 
  
We particularise \eqn{diffusivity} to IGWs and geostrophic flows using the dispersion relation \eqn{frequency} and the geostrophic balance satisfied by the velocity in $\Pi_{mn}$. It is natural to use spherical polar coordinates $(k,\phi,\theta)$ in $\bk$-space and a Fourier counterpart to $\Pi_{mn}$ in the form of the vortical flow kinetic energy spectrum $E(K_\mathrm{h},K_\mathrm{v})$, which we assume to be horizontally isotropic so that it only depends on the horizontal and vertical wavenumbers $K_\mathrm{h}$ and $K_\mathrm{v}$ (for clarity we systematically use lowercase symbols for coordinates in the IGW wavenumber space and uppercase symbols for coordinates in the flow wavenumber space). Computations detailed in Appendix \ref{app:diffusionEquation} then reduce \eqn{diffusion} to
\beq
\partial_t a = \frac{1}{k^{2}} \partial_{k} ( k^2 D_{kk} \partial_k a) +  \frac{D_{\phi \phi}}{k^2 \sin^2 \theta} \partial_{\phi\phi} a,
\lab{diffusionSpherical}
\eeq
under the further assumption of spatial homogeneity $\nabla_{\bx}a=0$. This makes it plain that there is no diffusion in the direction of $\theta$. Hence, $\theta$, or equivalently $\omega$, can be treated as a fixed parameter. The only non-zero components of the diffusivity tensor are given by
\begin{subequations}
\begin{align}
\lab{Dkk}
D_{kk} &= B k^3  \iint \displaylimits_{K_\mathrm{h}^2/K_\mathrm{v}^2>\tan^2{\theta}} \frac{K_\mathrm{v}^2}{K_\mathrm{h}} \left(\cot^2 \theta - \frac{K_\mathrm{v}^2}{K_\mathrm{h}^2} \right)^{1/2} E(K_\mathrm{h},K_\mathrm{v}) \, \d K_\mathrm{h} \d K_\mathrm{v},  \\ 
\frac{D_{\phi\phi}}{\sin^{2}\theta} &= B k^3    \iint \displaylimits_{K_\mathrm{h}^2/K_\mathrm{v}^2>\tan^2{\theta}} K_\mathrm{h} \left(\cot^2 \theta - \frac{K_\mathrm{v}^2}{K_\mathrm{h}^2} \right)^{3/2} E(K_\mathrm{h},K_\mathrm{v}) \, \d K_\mathrm{h} \d K_\mathrm{v},
\end{align}
\lab{diffusivitySpherical}
\end{subequations}
where 
\beq
B = \frac{\omega \sin^2 \theta}{4 \pi^3 (N^2-f^2)|\cos^5 \theta|}
\eeq
depends solely on $\theta$, $N$ and $f$.

Eqs.\ \eqn{diffusionSpherical}--\eqn{diffusivitySpherical} provide a full description of the diffusion of IGW on the constant-frequency cone in $\bk$-space for a turbulent flow of given energy spectrum. In the angular $\phi$-direction, this diffusion leads to an isotropisation of the wave field with rate $D_{\phi\phi}/(k^2 \sin^2 \theta)$. In the radial $k$-direction, the diffusion leads to a forward cascade of the wave energy to high wavenumbers where it is efficiently dissipated by viscous processes. Note that wave energy remains confined to one nappe of the cone corresponding to either upward- or downward-propagating IGWs. This is only an approximation; exchanges between upward- and downward-propagating waves do occur, but they are asymptotically small and not captured by the WKB approximation. In what follows, we concentrate on  radial diffusion by assuming wave statistics independent of $\phi$, $\partial_\phi a =0$, leaving the study of horizontal isotropisation for future work. 

\section{Initial-value problem} \label{sec:ivp}

For $a=a(k,t)$, we rewrite \eqn{diffusionSpherical} as
\beq
\partial_t e = \partial_{k} \left(Q k^5 \partial_k \left(k^{-2} e \right)\right), 
\lab{diffusionRadial}
\eeq
where we have introduced $e(k,t)=2\pi k^2 \, \sin \theta \, \omega a(k,t)$ and the $k$-independent parameter $Q=D_{kk}/k^3$. The function $e(k,t)$ is the IGW energy density in $k$, with $e(k,t) \, \d k$ the energy contained within the interval $[k,k+\d k]$.
We solve \eqn{diffusionRadial} with initial condition $e(k,0)=\delta(k-k_*)$ corresponding to the excitation of IGWs with a single wavenumber $k_*$. (The solution associated with arbitrary initial condition can be deduced by integration over $k_*$.) We show in Appendix \ref{app:ivpSolution} that
\beq
e(k,t)= \tfrac{1}{2} k_*^{-2} \int_0^\infty J_4(k^{-1/2} \lambda) J_4(k_*^{-1/2} \lambda) \e^{-Q\lambda^2t/4}  \lambda \, \d \lambda,
\lab{eBessel}
\eeq
where $J_4$ is a Bessel function of the first kind \citep{DLMF}.
The large-time behaviour of $e(k,t)$ is readily deduced as $e(k,t) \propto k^{-2} t^{-5}$ away from an asymptotically small neighbourhood of $k=0$ (see Appendix \ref{app:ivpSolution}). 
An inverse diffusion time scale can be read off from \eqn{eBessel} as $Q k_*$. Using \eqn{diffusivitySpherical} this can be written in the dimensionless form
\beq
\frac{Q k_*}{\omega} = \gamma \frac{N^2}{N^2-f^2} \frac{k_*}{K_{\mathrm{h}*} }  \mathrm{Ro^2},
\lab{timeScale}
\eeq
where $\gamma$ is a dimensionless `geometric' factor that depends only on $\theta$ and the shape (but not the magnitude) of the flow kinetic-energy spectrum  and $\mathrm{Ro} = K_{\mathrm{h}*} \av{|\bs{U}|^2}^{1/2}/f$ is a flow Rossby number. The typical horizontal and vertical inverse flow scales $K_{\mathrm{h}*}$ and $K_{\mathrm{v*}}$ are assumed to be related by  $K_{\mathrm{v}*} = N K_{\mathrm{h}*}/f $. Eq.\ \eqn{timeScale} captures the dependence of the diffusion time scale on the Rossby number and on the scale separation between IGWs and  flow. The diffusion approximation requires $Q k_*/\omega \ll 1$ in addition to the WKB conditions $k_*\sin \theta \gg K_{\mathrm{h}}$ and 
$k_*\cos \theta \gg K_{\mathrm{v}}$.

\begin{figure}
  \centering
 \begin{tabular}{cc}
   (a) & (b) \\
     \centering
 \includegraphics[trim={0 0.5cm 0cm 0cm},width=.47\linewidth]{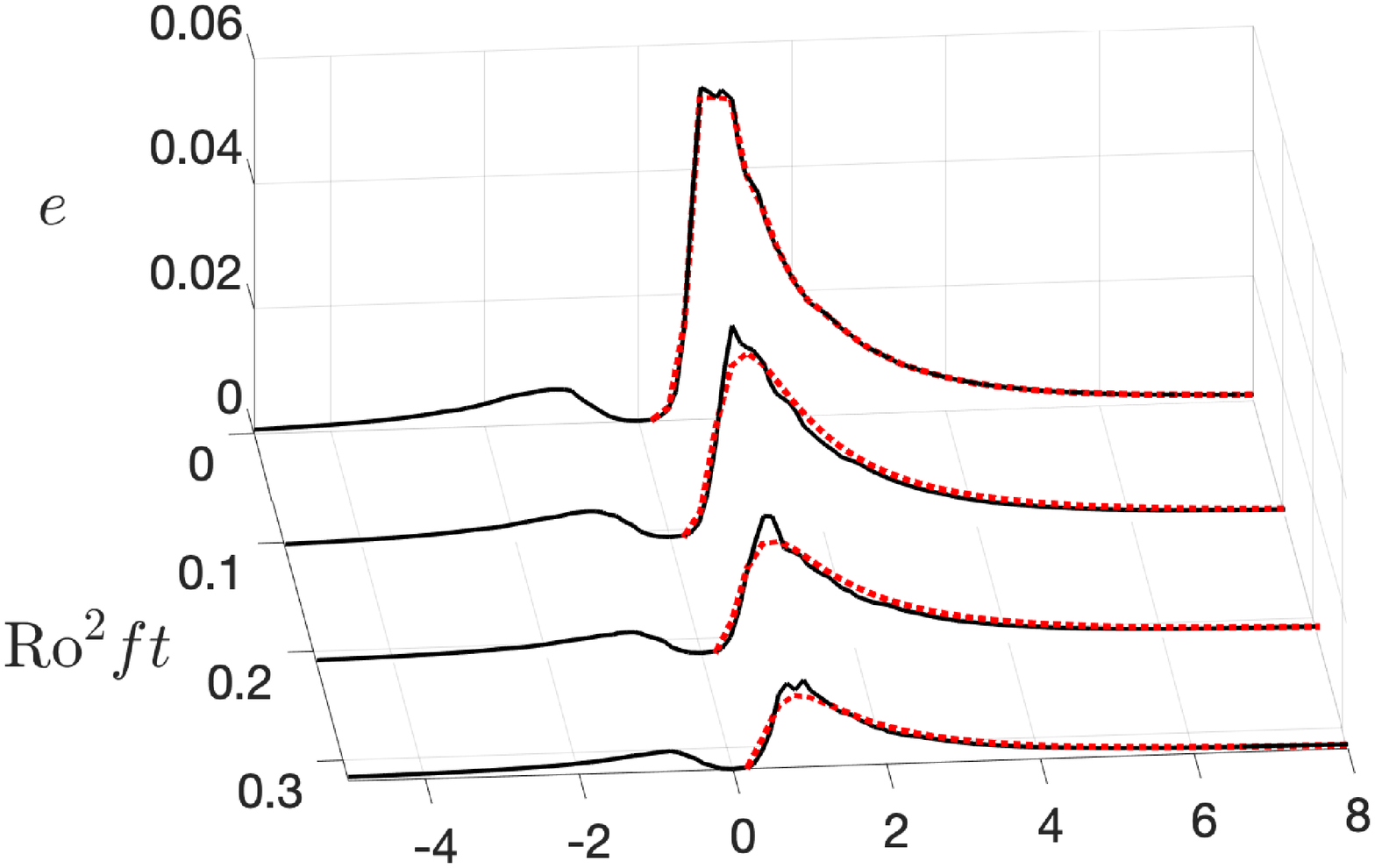}
& \includegraphics[trim={0 0.5cm 0cm 0cm},width=.47\linewidth]{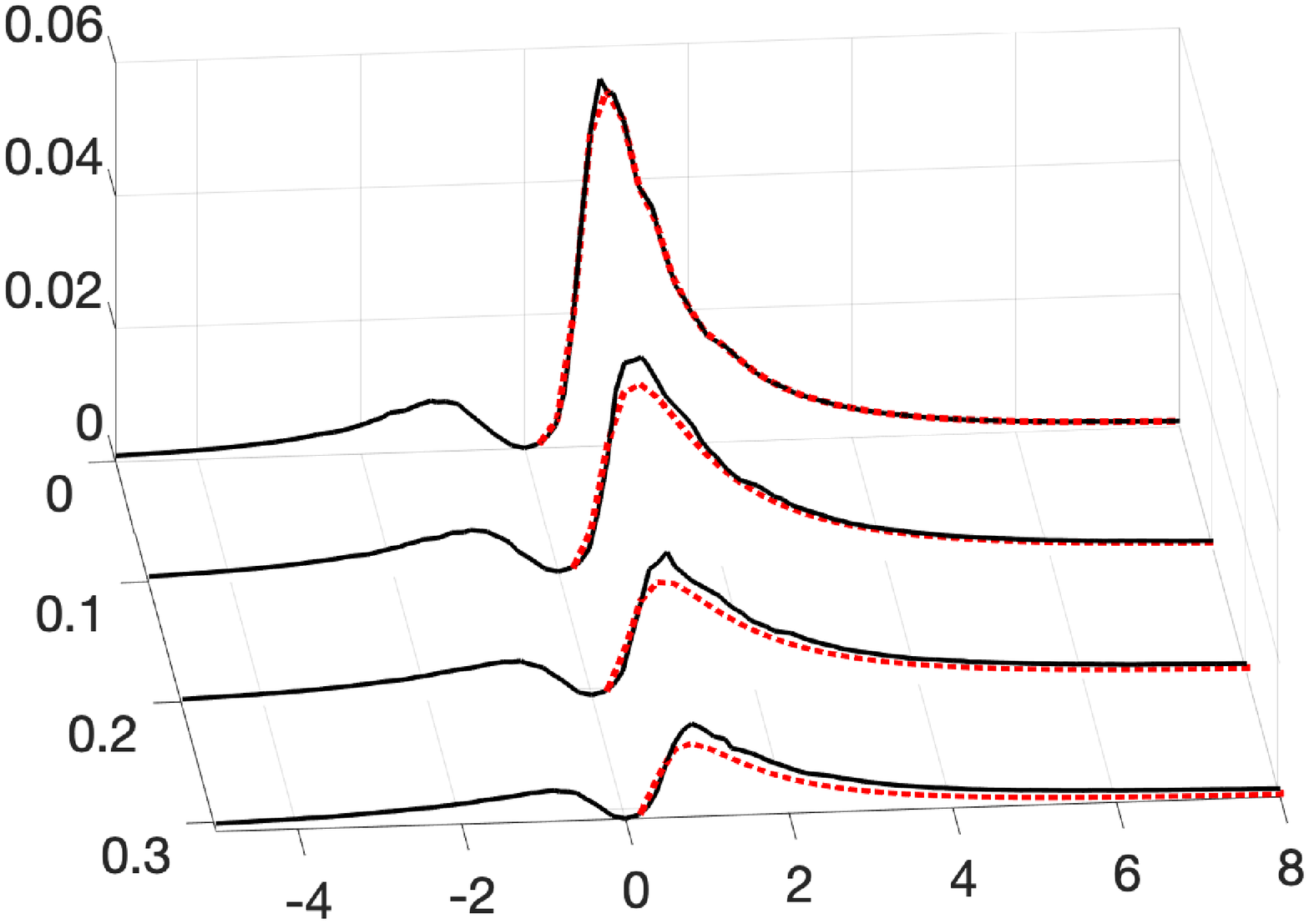} \\ 
   (c) & (d)   \\
      \centering
\includegraphics[trim={0 0.5cm 0cm 0.1cm},width=.47\linewidth]{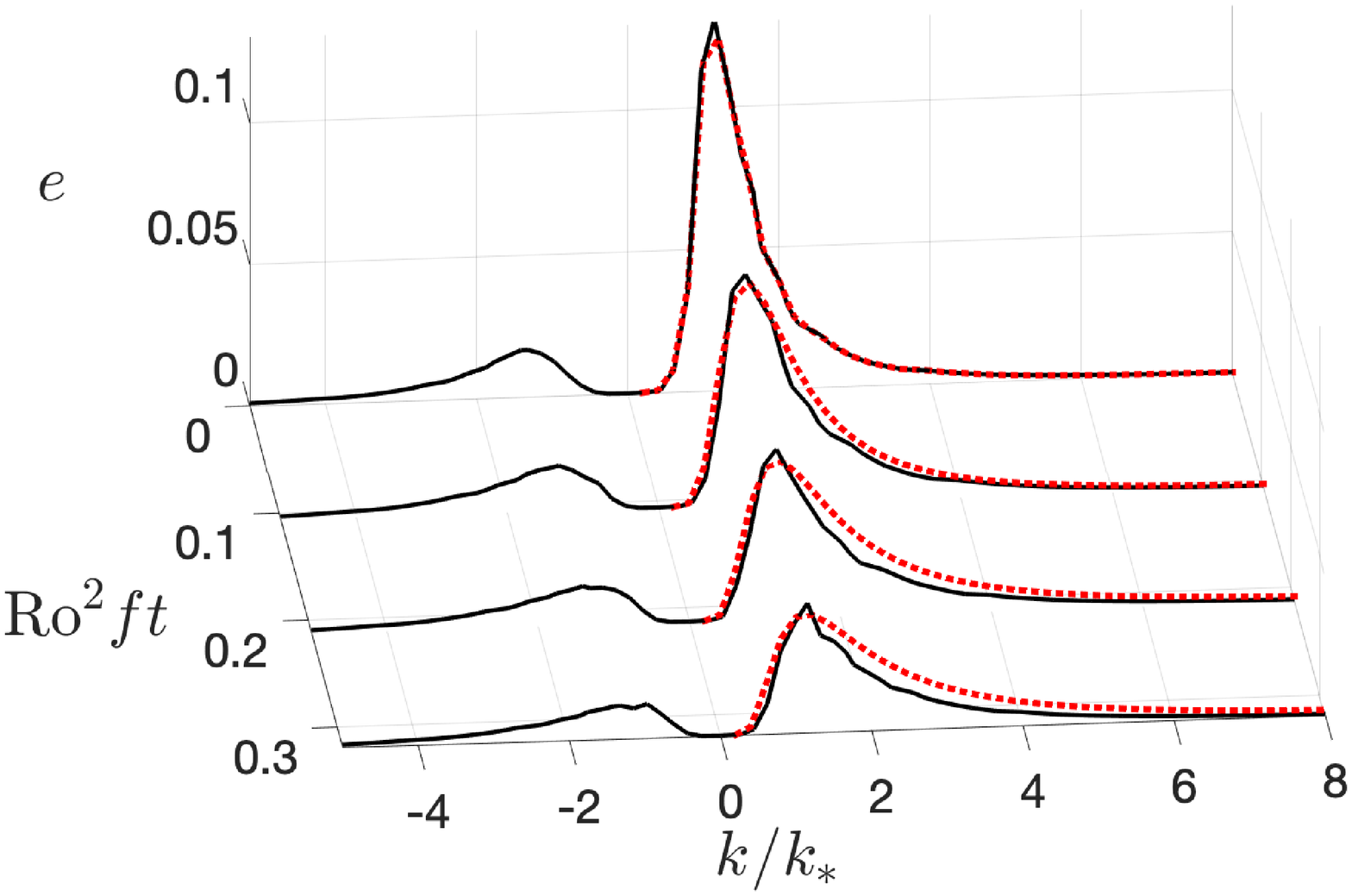}
& \includegraphics[trim={0 0.5cm 0cm 0.1cm},width=.47\linewidth]{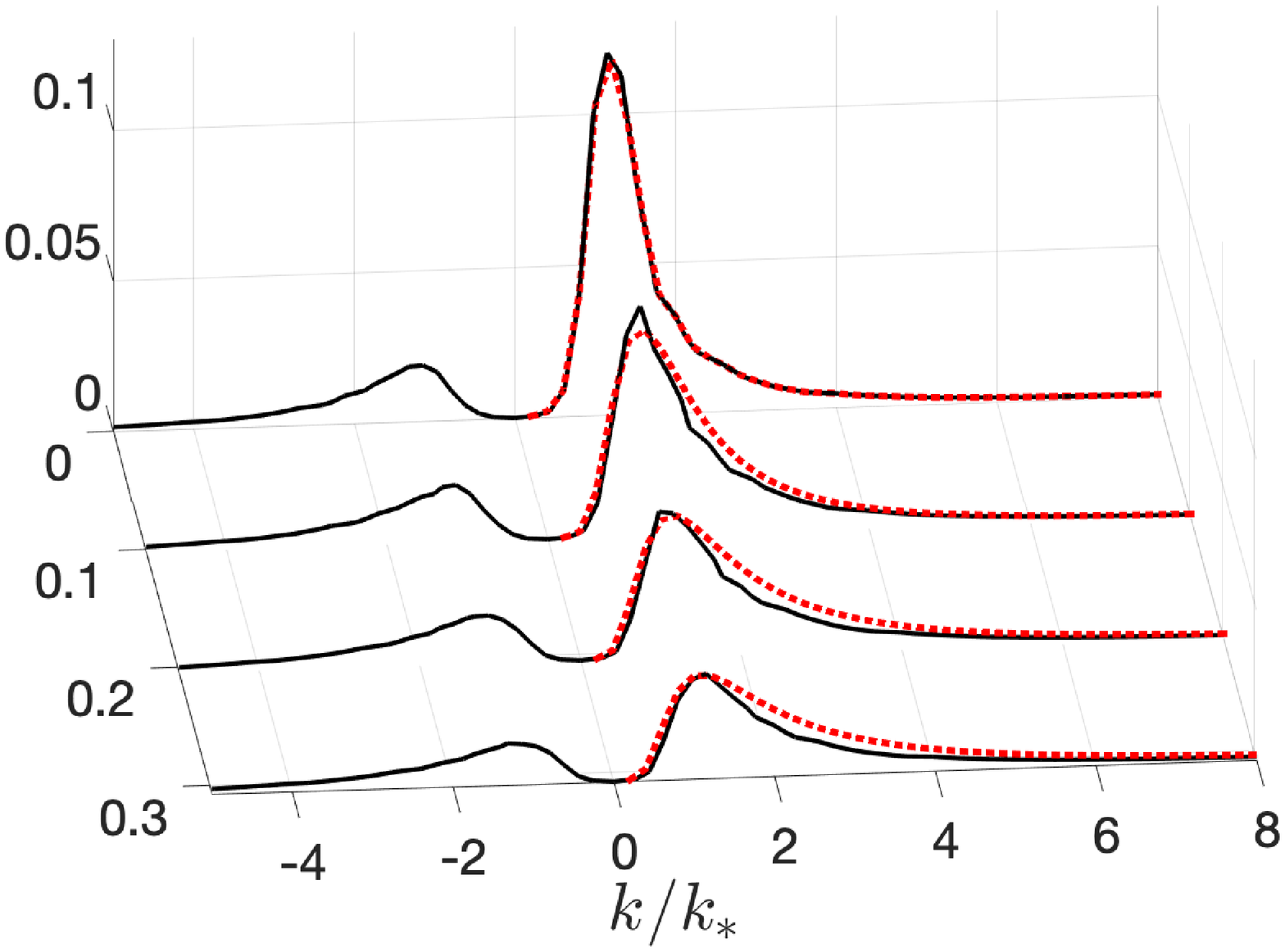} 
\end{tabular} 
\caption{Evolution of the IGW energy $e(k,t)$ in Boussinesq simulations (solid, black) and as predicted by the diffusion approximation (dotted, red) for (a) $\omega=2f, \, \mathrm{Ro}=0.057$, (b) $\omega=2f, \, \mathrm{Ro}=0.117$, (c) $\omega=3f, \, \mathrm{Ro}=0.057$ and (d) $\omega=3f, \, \mathrm{Ro}=0.117$. Conventionally, $k>0$ ($k<0$) corresponds to upward- (downward-)propagating IGWs.}
\label{fig:waterfalls}
\end{figure}

We verify the solution of \eqn{diffusionRadial} against simulations of the three-dimensional non-hydrostatic Boussinesq equations. 
These are solved using a code adapted from that in \citet{wait-bart06b} which relies on a de-aliased pseudospectral method and a third-order Adams--Bashforth scheme with timestep $0.015/f$. The triply-periodic domain, $(2 \pi)^3$ in the scale coordinates $(x,y,z'=Nz/f)$, is discretised uniformly with $768^3$ grid points. A hyperdissipation of the form $-\nu  (\partial_x^2+\partial_y^2 + \partial_{z'}^2)^4$, with $\nu = 2\times10^{-17}$,  is employed in the momentum and density equations.  We take $N/f=32$, a representative value of mid-depth ocean stratification. The initial condition is the superposition of a  turbulent flow, obtained by running a quasigeostrophic model to a statistically stationary state, and IGWs. The initial spectrum of the vortical flow peaks at $K_{\mathrm{h*}} \simeq 4$ and has an inertial subrange scaling approximately as $K_\mathrm{h}^{-3}$ and $K_\mathrm{v}^{-3}$.  This spectrum evolves slowly over the IGW-diffusion timescale, and an average is used to calculate $D_{kk}$ in \eqn{Dkk}, and hence $Q$ in \eqn{diffusionRadial}. 
We report experiments with the two  Rossby numbers $\mathrm{Ro}= K_{\mathrm{h}*} \av{|\bs{U}|^2}^{1/2}/f =0.057,\, 0.117$ (or $ \av{\zeta^2}^{1/2}/f=0.1,\, 0.2$ for the alternative Rossby numbers based on the vertical vorticity $\zeta$), and the two  IGW frequencies $\omega=2f, \, 3f$. 
Upward-propagating IGWs are initialised as a ring in $\bk$-space with $k_\mathrm{h*}=16$, $k_\mathrm{v}=\mathrm{\cot} \, \theta \, k_\mathrm{h}$, random phases, and an initial kinetic energy $\av{|\bs{u}|^{2}}/2 = 0.1 \av{|\bs{U}|^{2}}/2$. 
The IGW spectrum $e(k,t)$ is computed following the normal-mode decomposition of \citet{bart95}.

 \begin{figure}
  \centering
 \begin{tabular}{cc}
     \centering
\includegraphics[trim={0 0.5cm 0cm 0.1cm},width=.48\linewidth]{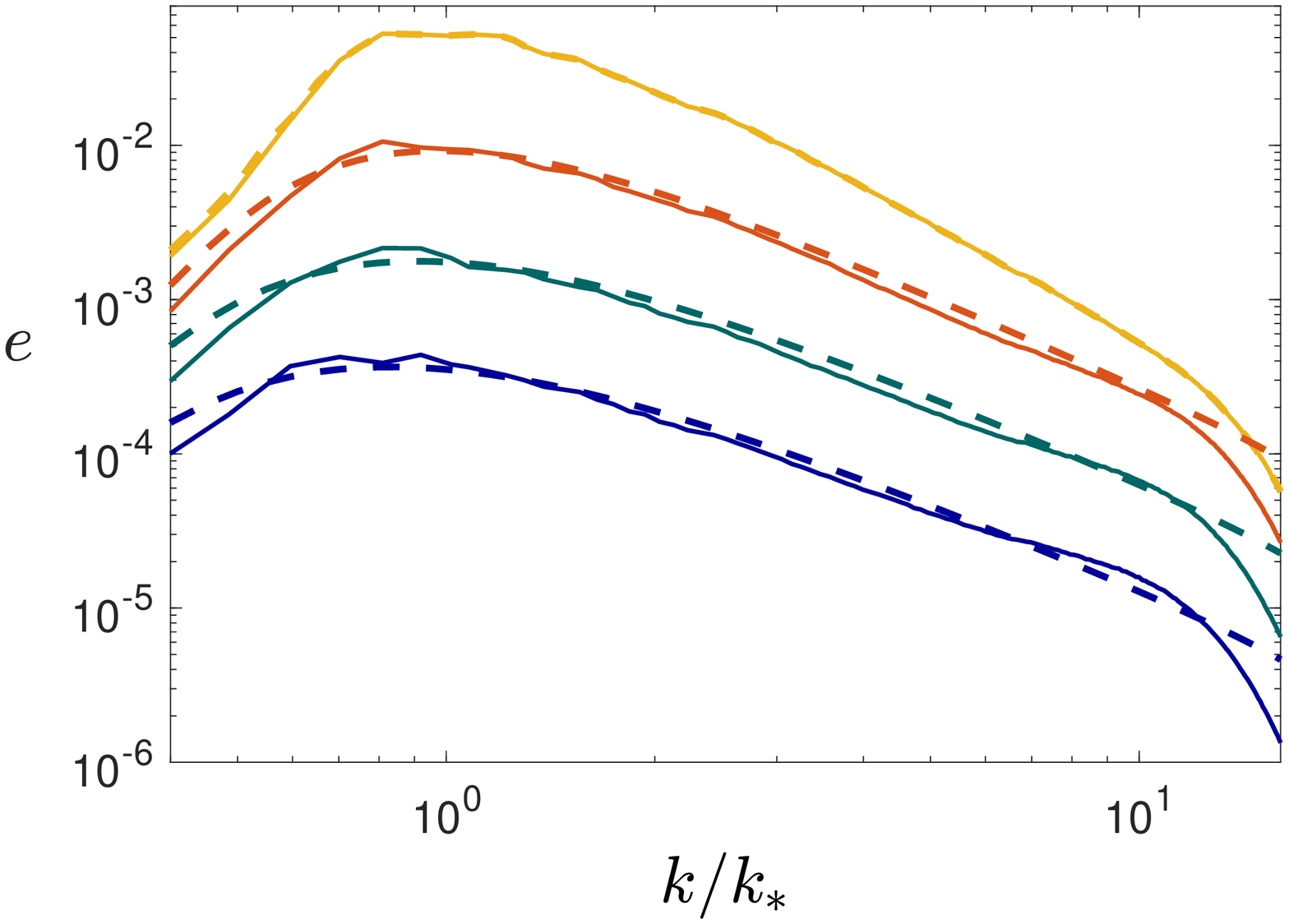}
& \includegraphics[trim={0 0.5cm 0cm 0.1cm},width=.48\linewidth]{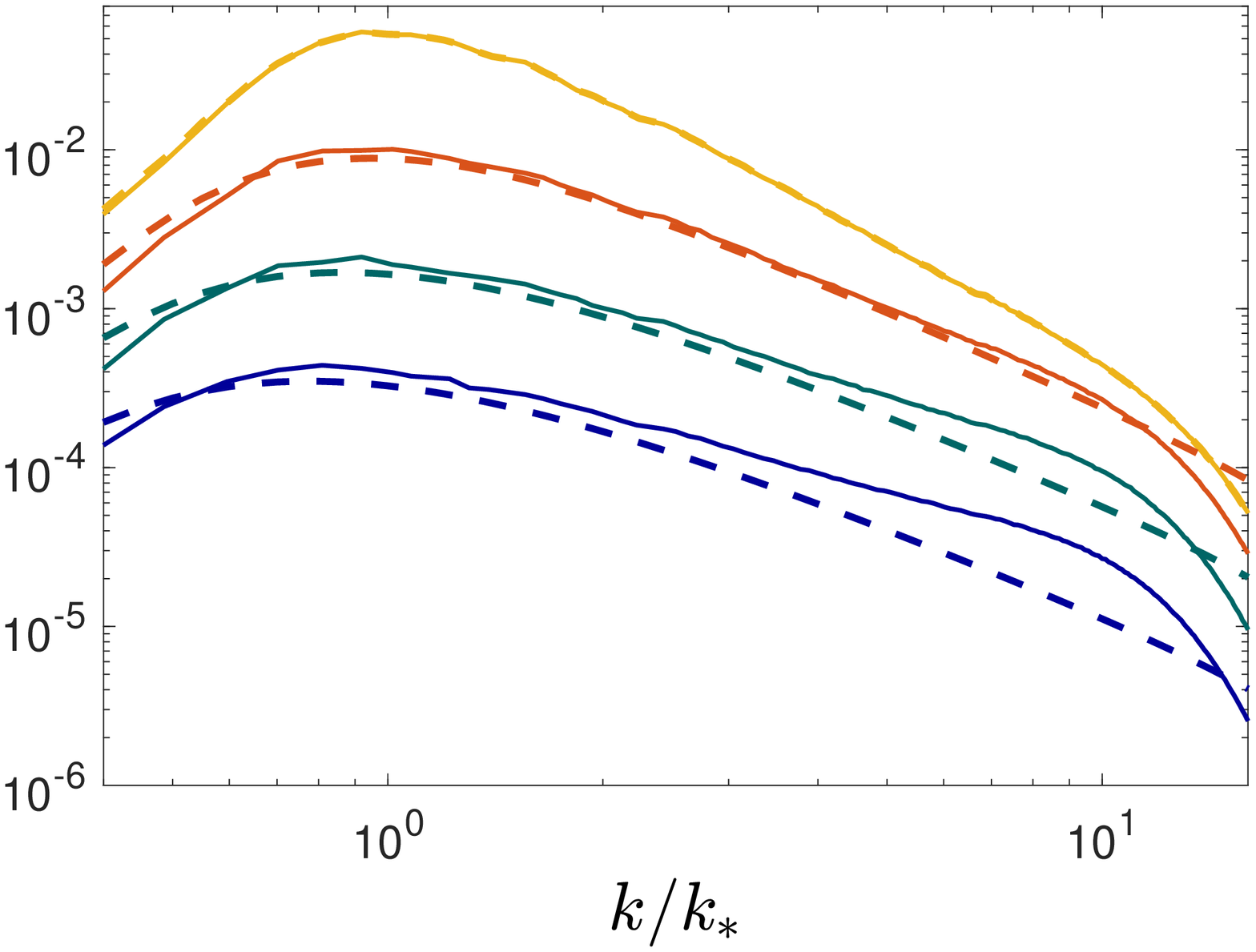} 
\end{tabular}  
\caption{Log-log representation of the IGW spectra in Figs.\ \ref{fig:waterfalls}(a) (left, $\omega=2f, \, \mathrm{Ro}=0.057$) and \ref{fig:waterfalls}(b) (right, $\omega=2f, \, \mathrm{Ro}=0.117$). The solid lines are the results of the Boussinesq simulation and the dashed lines the predictions of the diffusion approximation. The curves correspond to the times shown in Fig.\ \ref{fig:waterfalls}(a--b) and are successively shifted downward by half a decade for clarity.}
\label{fig:loglogs}
\end{figure}

Fig.\ \ref{fig:cones}, obtained for the lower $\mathrm{Ro}$ and $\omega = 3f$, illustrates the confinement of  wave energy on the constant-frequency cone, one of the keys to the validity of the diffusion approximation. The confinement is of course not perfect and some energy appears around the cones associated with the harmonic frequencies $2 \omega$ and $3 \omega$.
Fig.\ \ref{fig:waterfalls} shows the evolution of $e(k,t)$ for the four sets of values of $(\mathrm{Ro},\omega)$. The numerical results are compared with the predictions of the diffusion equation obtained by solving \eqn{diffusionRadial} initialised with the form of $e(k,t_\mathrm{a})$ extracted from the simulation after an adjustement time $t_\mathrm{a}>0$. This procedure accounts for the fact that the diffusion equation \eqn{diffusion} is only valid after an adjustment period, requiring $t_\mathrm{a} \gg (K_*|\bc|)^{-1}$, the time to traverse typical eddies at the IGW group speed \citep[cf.][\S5]{mull-et-al}. The agreement between the numerical simulation and the diffusion approximation is remarkable considering the complexity of the full Boussinesq dynamics and the moderate separation of scales between IGWs and flow. As the diffusion approximation predicts, the simulations with different Rossby numbers behave similarly when $t$ is scaled suitably. The decay is slower for $\omega=3f$ than $\omega=2f$, consistent with a decrease in $Q$ obtained when evaluating \eqn{Dkk}.
Scattering from upward-propagating to downward-propagating IGWs, neglected in the diffusion approximation, occurs; it is more substantial for the larger $\omega$ because the two nappes of the constant-frequency cones are closer together, facilitating energy transfers. Fig.\ \ref{fig:loglogs} displays the wave energy spectrum $e(k,t)$ obtained for $\omega = 2 f$ (top row of Fig.\ \ref{fig:waterfalls}) in log--log coordinates. It shows that the good agreement between numerical and predicted spectra extends to large wavenumbers for $\mathrm{Ro}=0.057$ but not for the larger Rossby, $\mathrm{Ro}=0.117$, at the later times. We note that the wave energy is then very small and may  be affected by a contribution associated with spontaneous generation \cite[cf.][]{kafi-bart}
 
%
%

\section{Forced response and observed ocean and atmosphere spectra} \label{sec:forced}

We now turn to the steady solution of \eqn{diffusionRadial} in the presence of a forcing of the form $\delta(k-k_*)$. Eq.\ \eqn{eBessel} admits two steady solutions: the no-flux solution $e(k) \propto k^2$ and the constant-flux solution $e(k) \propto k^{-2}$. Matching these at $k_*$ yields the steady spectrum
\beq
e(k) = \frac{1}{4Qk_*^2} \begin{cases}
(k/k_*)^2 & \textrm{for} \ \ 0< k < k_* \\
(k_*/k)^2 & \textrm{for} \ \  k > k_*
\lab{forcedSpectrum}
\end{cases}.
\eeq
Note that for IGWs with a single frequency and correspondingly a single angle $\theta_*$, the horizontal energy spectrum $e_\mathrm{h}(k_\mathrm{h}) $ satisfies the same power laws as $e(k)$ since
\beq 
e_\mathrm{h}(k_\mathrm{h}) = \iiint \delta( k \sin \theta-k_\mathrm{h}) \frac{e(k)\delta(\theta-\theta_*)}{2\pi k^2 \sin \theta}  \, \d \bk =  \csc \theta_* e(k_\mathrm{h} \csc \theta_*),
\eeq
using that the energy density in $\bk$-space is $e(k) \delta(\theta-\theta_*)/(2\pi  k^2 \sin\theta)$. 
Thus, \eqn{forcedSpectrum} implies a $k_\mathrm{h}^{-2}$ horizontal spectrum at large $k_\mathrm{h}$. This remains true for a superposition of IGWs with different frequencies, corresponding to an integration over $\theta_*$.

\begin{figure}
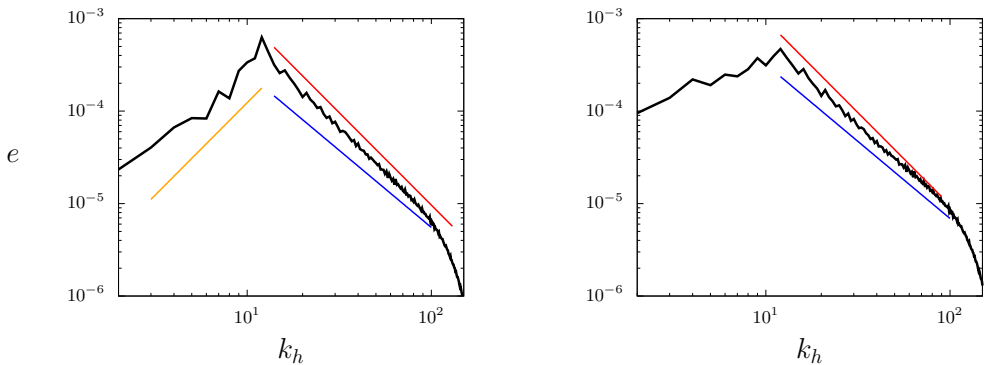

  \centering
 \begin{tabular}{cc}
 \resizebox{.5\linewidth}{!}{\input{fig4a.tex}}
 \resizebox{.5\linewidth}{!}{\input{fig4b.tex}}
 \end{tabular} 
  \caption{Stationary horizontal energy spectrum $e(k_\mathrm{h})$ for the forced simulations in \S \ref{sec:forced} with $\mathrm{Ro}=0.057$ (left) and $0.117$ (right). Straight lines indicate the power laws: \myredline   $\ k^{-2}$,  \myblueline $\ k^{-5/3}$  and  \myorangeline $\ k^{2}$.}
  \label{fig:forcedspc}
\end{figure}

We confirm the prediction \eqn{forcedSpectrum} by the simulation of the Boussinesq equations in the presence of forcing. In the simulation reported, all the specifications are the same as in \S\ref{sec:ivp} except for the initial condition, which is devoid of IGWs. Instead, an Ornstein--Uhlenbeck forcing with short correlation time (3 timesteps) is applied to the waves with $\omega = 2f$  \citep[see][]{wait17}.
%
%
The forcing amplitude is adjusted so that the wave energy is about $0.01$ of the vortical flow energy after reaching a stationary state. Fig.\ \ref{fig:forcedspc} shows the stationary spectra for $\mathrm{Ro}=0.057$ and  $0.117$. For the low Rossby number, the prediction \eqn{forcedSpectrum} is well borne out by the simulation results with a clean $k_\mathrm{h}^{-2}$ spectrum spanning nearly a decade from the forcing scale down to dissipation. For the high Rossby number, the spectrum shallows a little from wavenumber $40$ or so to take a shape more consistent with $k_\mathrm{h}^{-5/3}$. Two mechanisms can be invoked to explain this shallowing: the Doppler term is not weak compared with the intrinsic IGW frequency, invalidating the diffusion approximation, or nonlinear wave--wave interactions become significant. We can roughly estimate the wavenumbers at which each of these mechanisms is important as 
\beq
k_\mathrm{h} \sim \frac{\omega}{\langle|\bU|^2 \rangle^{1/2}} \sim \frac{K_{\mathrm{h}*}}{\mathrm{Ro}}  \inter{and} k_\mathrm{h} \sim \frac{K_{\mathrm{h}*}}{\mathrm{Ro}} \frac{\langle{|\bU|^2 \rangle^{1/2}}}{\langle|\bu|^2 \rangle^{1/2}},
\specialnumber{a,b}
\lab{khkh}
\eeq
corresponding to order-one Rossby numbers based on the wave lengthscale $k_{\mathrm{h}}^{-1}$ and on the root-mean-square velocity of, respectively, the vortical flow and the IGWs. For the simulation with $\mathrm{Ro}=0.117$, these wavenumbers are about 40 and 400, suggesting that the shallowing of the spectrum is associated with the breakdown of the assumption of weak Doppler shift.

The prediction of a $k_\mathrm{h}^{-2}$ spectrum is significant in view of the ubiquity of this scaling in ocean and atmosphere observations.
In the ocean, kinetic energy spectra show a $k_\mathrm{h}^{-2}$ dependence in the submesoscale range, say below 20 km, in regions of high mesoscale activity and in a larger range, below 200 km, in less active regions (see \citet{cali-ferr}  for a comprehensive discussion).  
Recent analyses by \citet{buhl-et-al14} and \citet{roch-et-al16} which separate the contribution of IGWs from that of  geostrophic motion indicate that the IGW part of the spectrum follows a $k_\mathrm{h}^{-2}$ scaling in almost the entirety of its range. Our results above suggest that this may result from IGW energy diffusion by the geostrophic flow. Scales below 10 km or so are the realm of the \citet{garr-munk}
 spectrum, also associated with a $k_\mathrm{h}^{-2}$ dependence. While this spectrum is generally attributed to wave--wave interactions \citep[e.g.][]{mull-et-al,lvov-et-al}, interactions with the geostrophic flow may play a significant role, dominating for wavenumbers much smaller than (\ref{eqn:khkh}b). We emphasise that theories based on linear IGWs, be it the diffusion approximation of this paper or a more general theory accounting for strong Doppler shift, cannot predict the level of IGW spectrum nor its frequency content since both are determined by the forcing.

\begin{figure}
  \centering
 \resizebox{.6\linewidth}{!}{\input{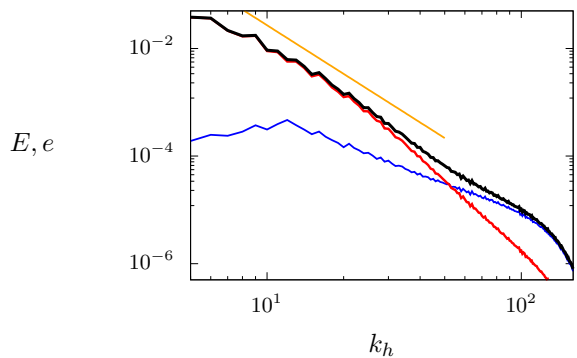}}
  \caption{Stationary horizontal energy spectrum for the forced simulations in  \S \ref{sec:forced} with $\mathrm{Ro}=0.117$: \myblackline \, total energy, \myredline \, vortical energy $E$,  \myblueline \, IGW energy $e$ (same as on the right panel of Fig.\ \ref{fig:forcedspc}), and \myorangeline \, $k_\mathrm{h}^{-3}$ power law.}
  \label{fig:wavebalancespc}
\end{figure}

In the atmosphere, similarly, there is a broad range of scales, from 500 km to 10 km, where the energy spectrum scales approximately as $k_{\mathrm{h}}^{-2}$. This is the shallow, mesoscale part of the celebrated \citet{nast-gage} spectrum, which is traditionally interpreted as a $k_{\mathrm{h}}^{-5/3}$ spectrum but is also consistent with $k_{\mathrm{h}}^{-2}$. There is ongoing debate about the nature of this part of the spectrum: \citet{cali-et-al14,cali-et-al16} attribute it to nearly linear IGWs on the basis of their separation between IGWs and geostrophic motion, but this interpretation is controversial (see \citet{li-linb} for a recent critique). 
\citet{cali-et-al16} note that `the wave interpretation is \ldots\, not inconsistent with the observed power-law spectra \dots\, but an explanation for the spectral shape is so far missing'. Our results provide a possible explanation. 

The total spectrum in the high-$\mathrm{Ro}$ simulation, shown in Fig.\ \ref{fig:wavebalancespc}, is reminiscent of atmospheric observations, with a $k_\mathrm{h}^{-3}$ range at large scales associated with the vortical flow, a $k_\mathrm{h}^{-2}$ range associated with nearly linear IGWs at intermediate scales, and a  further shallowing at small scales (best seen in Fig.\ \ref{fig:forcedspc}, right panel). While the diffusion approximation explains the $k_\mathrm{h}^{-2}$ range in our simulations, a degree of caution is required to draw a similar conclusion for the atmospheric spectrum since some of the underlying assumptions -- weak flow with homogeneous  statistics and relevance of the equilibrium spectrum in particular -- are questionable.




\section{Discussion}
 
This paper examines the impact of a turbulent geostrophic flow on the statistics of small-amplitude IGWs. This impact has  received less attention than that paid to wave--wave interactions. Yet the timescale found for a substantial effect of the geostrophic flow, of the order of $0.1 \, \mathrm{Ro}^{-2} f^{-1}$ (see Fig.\ \ref{fig:waterfalls})  corresponding to tens of days for ocean parameters, is similar to that of the fastest wave--wave interaction process (parametric subharmonic instability of internal tides at the critical latitude $29^\circ$, \citet{mack-wint}). This confirms the conclusions of \citet{ward-dewa} and \citet{savv-vann} that scattering by the flow dominates over wave--wave interactions in many ocean circumstances. A similar conclusion has been drawn from numerical simulations \citep{wait-bart06a}. 

The present paper focuses on the diffusive regime of IGW scattering that arises for weak flows and small-scale, linear IGWs. A remarkable feature of this regime is the prediction of a $k^{-2}_\mathrm{h}$ energy spectrum consistent with  observations in both the ocean and atmosphere. 
When the assumption of small scales is relaxed, the wave energy obeys a kinetic equation generalising the equations obtained by \citet{dani-v16} and \citet{savv-vann} in the case of inertial waves and IGWs in a barotropic flow. The kinetic equation captures the transfer of energy between upward and downward-propagating IGWs which is negligible in the diffusive regime. The derivation and analysis of this equation are the subject of ongoing work. When the assumption of weak flow is relaxed, as required for wavenumbers not small compared with (\ref{eqn:khkh}a), IGWs are in the eikonal regime considered by \citet{heyn-pomp} in the context of wave--wave interactions \cite[see also][\S5]{mull-et-al}. It would be desirable to study the scattering by geostrophic flow in this regime. We conclude by noting that the consistency between predicted and observed spectral slopes is only indicative: further investigations are needed to establish the importance of IGW scattering in determining oceanic and atmospheric spectra.


\medskip
 
\noindent 
\textbf{Acknowledgements.} This research is funded by the UK Natural Environment Research Council under the NSFGEO-NERC programme (grant NE/R006652/1). M.A.C.S was supported by the Maxwell Institute Graduate School in Analysis and its Applications, funded by the UK Engineering and Physical Sciences Research Council (grant EP/L016508/01), the Scottish Funding Council, Heriot-Watt University and the University of Edinburgh. This work used the ARCHER UK National Supercomputing Service.
 
\appendix

\section{Derivation of the diffusion equation and of its solution} \label{app:diffusionEquation}

\subsection{General wave systems}

We introduce a small parameter $\eps \ll 1$ in the action conservation \eqn{action} by writing the frequency as $\Omega = \omega + \eps \bU \cdot \bk$, indicating that the velocity field is weak enough for the intrinsic frequency to dominate over the Doppler shift.
Defining slow time and spatial scales by $T=\eps^2 t$ and  $\bX=\eps^2 \bx$, we substitute the expansion
\beq
a=a^{(0)}(\bX,\bk,T) + \eps a^{(1)}(\bx,\bX,\bk,t,T) + \cdots
\eeq
into \eqn{action}. The first non-trivial equation appears at $O(\eps)$ and is given by
\beq
\partial_t a^{(1)} + c_i \partial_{x_i} a^{(1)} = k_m \partial_{x_i} U_m  \, \partial_{k_i} a^{(0)},
\eeq
using Cartesian components and implied summation. 
Assuming that the velocity field varies on the slow time scale only, the solution is given by
\beq
a^{(1)}(\bx,\bX,\bk,t,T) = k_m \int_0^t \partial_{x_j} U_m(\bx-\bc s,T) \, \d s  \, \partial_{k_j}  a^{(0)}.
\label{a1}
\eeq
Averaging the next-order equation to eliminate the terms containing $a^{(2)}$, we find
\beq
\partial_T a^{(0)} + c_i \partial_{X_i} a^{(0)} = k_n \langle \partial_{x_i} U_n  \partial_{k_i} a^{(1)} \rangle,  
\eeq
since $\langle U_i \partial_{x_i} a^{(1)} \rangle =\langle \partial_{x_i}( U_i  a^{(1)} )\rangle=0$ using incompressibility and spatial homogeneity.  Substituting the limit of \eqref{a1} as $t \to \infty$ as appropriate for the slow dynamics, we obtain the diffusion equation
\beq
\partial_T a^{(0)} + c_i \partial_{X_i} a^{(0)} =  \partial_{k_i} \left(\D_{ij} \partial_{k_j} a^{(0)} \right) 
\lab{diffapp}
\eeq
with the diffusivity
\beq
\D_{ij} = - k_m k_n \int_0^\infty \langle \partial_{x_i} U_n(\bx) \partial_{x_j} U_m(\bx-\bc s) \rangle \, \d s. 
\lab{diffuapp}
\eeq
This can be written as \eqn{diffusivity} in terms of the correlation tensor $\Pi_{mn}$ or, alternatively, as
\beq
\D_{ij} = \frac{k_m k_n}{8 \pi^2} \int_{\mathbb{R}^3} K_i K_j \hat \Pi_{mn}(\bK) \delta(\bK \cdot \bc) \, \d \bK,
\lab{diffufou}
\eeq 
in terms of the Fourier transform $\hat \Pi_{mn}$ of $\Pi_{mn}$. 
The diffusive approximation \eqn{diffapp} is standard for Hamiltonian systems with weak random perturbation and has been obtained in a variety of contexts (e.g., \citet{mcco-bret} for wave--wave interactions). The formal derivation above follows \citet[][\S4.2]{bal-et-al} 
who also discuss its rigorous justification.
%

\subsection{IGWs in quasigeostrophic flow} 

We particularise \eqn{diffufou} to the IGW dispersion relation \eqn{frequency} and a velocity field of the form $\bU=(-\partial_{x_2} \psi, \partial_{x_1} \psi,0)$ with $\psi$ the geostrophic streamfunction. 
We use the spherical polar coordinates $(k,\theta,\phi)$ for $\bk$, with $\be_k$, $\be_\theta$ and $\be_\phi$ the corresponding unit vectors, and express the group velocity as
\beq
\bc(\bk)= \frac{(N^2-f^2) \cos \theta \sin \theta}{k \omega} \, \be_\theta.
\eeq 
The diffusivity can be written in the basis $(\be_k,\be_\theta,\be_\phi)$ as
\beq
\DD= \D_{kk} \, \be_k \otimes \be_k + \D_{k \phi} (\be_k \otimes \be_\phi + \be_k \otimes \be_\phi) + \D_{\phi \phi} \, \be_\phi \otimes \be_\phi,
\lab{diffuspherical}
\eeq
where $\D_{kk}=\be_k \cdot \DD \cdot \be_k$, $\D_{k\phi}=\be_k \cdot \DD \cdot \be_\phi$, $\D_{\phi\phi}=\be_\phi \cdot \DD \cdot \be_\phi$, and we have made use of the fact that $\DD \cdot \be_\theta \propto \DD \cdot \bc = 0$ to eliminate all components along $\be_\theta$. 

With $\Theta$ and $\Phi$ the polar and azimuthal angles of the flow wavevector $\bK$,  we have
\begin{align}
 \bK &= K  (\sin{\Theta}\sin{\theta} \cos{\gamma} + \cos{\Theta}\cos{\theta}) \, \be_k  \nonumber \\
    &+ K \sin{\Theta}\sin{\gamma} \, \be_\phi 
    + K (\sin{\Theta}\cos{\theta} \cos{\gamma} - \cos{\Theta}\sin{\theta}) \, \be_\theta,
    \label{Knewcord}
\end{align}
where $\gamma=\Phi-\phi$.  Hence the delta function in \eqn{diffufou} can be written as
\beq
\delta(\bK \cdot \bc)=\frac{k \omega \left( \delta(\gamma-\gamma_*) + \delta(\gamma+\gamma_*) \right)}{K(N^2-f^2)\sin \Theta \sin \theta \cos^2 \theta \sin \gamma_*},
\lab{dirac}
\eeq
where $0 \le \gamma_* = \cos^{-1}(\tan \theta/\tan \Theta) \le \pi$. We also note that
\beq
k_m k_n \hat \Pi_{mn} = (k_1 K_2 - k_2 K_1)^2 \langle \hat \psi(\bK) \hat \psi(-\bK) \rangle = 2 k^2 \sin^2 \theta \sin^2 \gamma E(\bK),
\lab{aaa}
\eeq
where $E(\bK)=K^2 \sin^2 \Theta \langle \hat \psi(\bK) \hat \psi(-\bK) \rangle/2$ is the flow kinetic energy spectrum. We now introduce \eqn{dirac}--\eqn{aaa} into \eqn{diffufou} projected onto $\be_k$ and $\be_\phi$ to compute the components of $\DD$ in \eqn{diffuspherical}. Assuming that the flow is isotropic in the horizontal so that $E(\bK)$ is independent of $\gamma$, we obtain after some simplifications
\begin{subequations}
\begin{align}
D_{kk} &= \frac{k^3 \omega \sin^2{\theta}}{2\pi^2(N^2 - f^2)|\cos^5{\theta}|} \int_{-\infty}^\infty \int_{-\theta}^{\pi-\theta}
     K^3 \cos^2 \Theta (\cot^2{\theta}-\cot^2{\Theta})^{1/2} E(\bK) \ \d K \d \Theta, \\
D_{\phi\phi} &= \frac{k^3 \omega \sin^4{\theta}}{2\pi^2(N^2 - f^2)|\cos^5{\theta}|} \int_{-\infty}^\infty \int_{-\theta}^{\pi-\theta}
     K^3 \sin^2 \Theta (\cot^2{\theta}-\cot^2{\Theta})^{3/2} E(\bK) \ \d K \d \Theta,
\end{align} 
\end{subequations}
and $D_{k\phi}=0$. The form \eqn{diffusivitySpherical} follows by replacing the kinetic-energy spectrum $E(\bs{K})$ by its two-dimensional counterpart $E(K_\mathrm{h},K_\mathrm{v})=2\pi K_\mathrm{h} E(\bs{K})$ and changing the integration variables from $(K,\Theta)$ to $(K_\mathrm{h},K_\mathrm{v})$, with $K \, \d K \d \Theta=\d K_\mathrm{h} \d K_\mathrm{v}$. 

\subsection{Solution of \eqn{diffusionRadial} and its long-time approximation} \label{app:ivpSolution}

Introducing a solution of the separable form
$e(k,t) = \e^{-Q \lambda^2 t/4} f(k,\lambda)$, with $\lambda \ge 0$ a spectral parameter,
 into \eqn{diffusionRadial}
leads to 
\beq
k^3 f'' + k^2 f' + 4 \left( {\lambda^2}/{16}-k \right) f = 0,
\eeq
where the prime denotes derivative with respect to $k$. Solutions bounded as $k \to 0$ are proportional to the Bessel function $J_4(\lambda/\sqrt{k})$. The general solution of \eqn{diffusionRadial} follows as 
\beq
e(k,t) = \int_0^\infty A(\lambda) J_4(\lambda/\sqrt{k}) \e^{-Q \lambda^2 t/4} \, \d \lambda,
\eeq
for an arbitrary function $A(\lambda)$. Imposing the initial condition $e(k,0)=\delta(k-k_*)$ yields  \eqn{eBessel} on using the Bessel-function expansion of $\delta(k-k_*)$ \citep[][Eq.\ 1.17.13]{DLMF}.

For large $t$, the integral in \eqn{eBessel} is dominated by a neighbourhood of $\lambda=0$. The Bessel functions $J_4$ can therefore by replaced by their small-argument approximation,  $J_4(z) \sim z^4/16$ as $z \to 0$ \citep[][Eq.\ 10.2.1]{DLMF}, leading to
\beq
e(k,t) \propto k^{-2} \int_0^\infty \lambda^9 \, \e^{-Q \lambda^2 t/4} \, \d \lambda \propto k^{-2} t^{-5}.
\eeq

\bibliographystyle{jfm}
\bibliography{mybib}

\end{document}